\documentclass[a4paper,11pt]{article}
\usepackage{pos}
\usepackage{booktabs}
\usepackage{graphicx}
\usepackage{caption}
\usepackage{subcaption}
\usepackage{feynmp} 
\renewcommand{\logo}{\relax}

\title{Pontecorvo Reactions}

\author*[a,b]{Luca Venturelli }

 \onbehalf{on behalf of the ASACUSA collaboration}

\affiliation[a]{Dipartimento di Ingegneria dell’Informazione, Università degli Studi di Brescia\\
  via Branze 38, I-25123 Brescia, Italy}

\affiliation[b]{Istituto Nazionale di Fisica Nucleare, Sezione di Pavia,\\
via Bassi, I-27100 Pavia, Italy}

\emailAdd{luca.venturelli@unibs.it}

\abstract{
Pontecorvo reactions are rare antinucleon annihilation processes that are forbidden on free nucleons but allowed on nucleons bound within nuclei. The interest in studying this phenomenon lies in its potential to provide insights into the annihilation mechanism and, particularly, the short-distance dynamics between nucleons within the nucleus. 
Some measurements were performed in the past at CERN's Low Energy Antiproton Ring (LEAR) using antiprotons annihilating on a deuterium target. However, no data exist for targets consisting of three nucleons, such as $^3\text{He}$ or $^3\text{H}$.
The measurement of the rate of the process
$\overline{p} \, ^3\text{He} \rightarrow p + n$ 
would allow for distinguishing between different theoretical models whose predictions vary by 1-2 orders of magnitude. 
The ASACUSA collaboration is studying the feasibility of performing this measurement at CERN's ELENA-AD. 
A preliminary design of a simple measurement apparatus, utilizing plastic scintillators and degrader layers, is presented, together with Monte Carlo simulations assessing its efficiency in measuring the branching ratios of the aforementioned reaction and rejecting background from more probable typical antiproton annihilations in the target.
}

\FullConference{%
  EXA/LEAP2024,\\
  26-30 August 2024\\
  Austrian Academy of Sciences, Vienna, Austria
}

\begin{document}
\maketitle

\section{Introduction}

The scientific activities at the AD/ELENA facility at CERN focus on advanced experiments with antiprotons to test the fundamental laws of physics~\cite{Hori2013}. Key objectives include testing the Weak Equivalence Principle with antihydrogen in Earth’s gravity~\cite{Anderson2023,Kellerbauer2008,Perez2015} and verifying CPT invariance through precision spectroscopy techniques. These include laser spectroscopy on antihydrogen~\cite{Ahmadi2018a,Ahmadi2018b,ALPHA2020} and antiprotonic helium~\cite{Hori2011,Hori2016}, hyperfine microwave spectroscopy of antihydrogen~\cite{Ahmadi_2017,Malbrunot_2018}, and single-particle quantum transition spectroscopy~\cite{BASE_2015}.
%
These studies aim to address critical questions about the nature of the Universe, such as the origin of the observed unbalance between matter and antimatter.

AD/ELENA also hosts nuclear physics activities, studying antiproton-nucleus interactions, including exotic states like antiprotonic atoms~\cite{Doser2022,Zurlo_2006} and annihilation 
processes~\cite{Bianconi,Aghai_Khozani_2018,Aghai_Khozani_2021,Amsler_2024,PUMA_2022}, offering insights into nuclear structure and fundamental forces.

The annihilation of an antinucleon in matter is a process in which the antinucleon usually interacts with a single nucleon in the target nucleus~\cite{Amsler_LEAR}. However, processes in which multiple nucleons participate in the annihilation of the antinucleon are also possible.
Pontecorvo reactions are rare antinucleon annihilations that are forbidden on a free nucleons but permitted on nucleons bound within nuclei. The Italian physicist Bruno Pontecorvo suggested their existence in 1956~\cite{Pontecorvo_1956}, just a few months after the discovery of the antiproton at the Bevatron.
A typical example is the annihilation of an antiproton on a deuteron, resulting in the production of one meson and one nucleon (or a heavier baryon), such as 
$\;\overline{p} d \rightarrow \pi^- p \; $ and $\;\overline{p} d \rightarrow K^0 \Lambda$.
In Pontecorvo reactions, energy-momentum conservation is satisfied by allowing the process to proceed with the involvement of at least two nucleons.
Conversely, annihilation with a 'quasi-free nucleon' in a deuteron, where the other nucleon acts as a 'spectator,' necessitates producing at least two mesons.

The Pontecorvo reactions have been studied both theoretically and experimentally (for short reviews, see~\cite{Amsler_LEAR,Donnachie_2004}) .
The interest in this phenomenon lies in its potential to provide insights into the annihilation mechanism and particularly the short-distance dynamics between nucleons within the nucleus. 
However, making predictions based on first principles is challenging. Two main distinct approaches are used.
The first is a two-step process ("rescattering model")~\cite{Hernandez_Oset_1989,Kondratyuk_Sapozhnikov_1989,Kondratyuk_Guaraldo_1991,Kudryavtsev_Tarasov_1992,Kharzeev_Nichitiu_Sapozhnikov_1992,Kudryavtsev_Tarasov_1993}, while the second is a statistical method ("fireball" or "bag model")~\cite{Rafelski_1980,Rafelski_1988,Cugnon_Vandermeulen_1984,Cugnon_Vandermeulen_1989}. 

Focusing on the deuteron as the target, in the two-step model, it is initially hypothesized that the antinucleon annihilates with a nucleon, producing two mesons, one of which is absorbed by the other nucleon, resulting in the formation of a nucleon or a baryonic resonance.
For example: \(\overline{p} d \rightarrow \pi^- (\pi^+ n) \rightarrow \pi^- p\) or \(\overline{p} d \rightarrow \pi^- (\pi^+ n) \rightarrow \pi^- \Delta^+\).

In the statistical model, the 3 anti-quarks of the antinucleon and the 6 quarks of the 2 nucleons of the deuteron combine to form a compound system ("bag" or "fireball"), which, through quark rearrangements and annihilations, evolves into the final state.

Before the advent of LEAR, knowledge of these reactions was limited to rare observations, such as six bubble chamber events of the reaction \(\overline{p} d \rightarrow \pi^- p\)~\cite{Bizzarri_1969}.
Systematic exploration became feasible with experiments at LEAR, including ASTERIX, OBELIX, and Crystal Barrel, which utilized 
stopped antiprotons in gaseous and liquid deuterium to measure rare branching ratios~\cite{ASTERIX_1989,Denisov_1999,Amsler_1995,Gorchakov_2002,Ableev_1994,Amsler_B352_1995,Abele_1999}.
The LEAR experiments revealed small branching ratios (on the order of 10$^{-6}$ to 10$^{-5}$) for various channels, as reported in Table~\ref{tab:Pontecorvo}.
More recent measurements of specific channels have confirmed these results~\cite{Chiba1997}.

\begin{table}[htb]
\begin{center}
\begin{tabular}{ l l l l}
\hline
Reaction & \multicolumn{2}{c}{Branching ratio}&  Experiment\\
\hline
$\bar{p}d \to \pi^-p$ & 1.46 $\pm 0.08 $ & $\times 10^{-5}$ & \scriptsize{OBELIX (in gas)}~\cite{Denisov_1999}\\
$\bar{p}d \to \pi^0 n$ & 7.02 $\pm 0.72 $ &$\times 10^{-6}$ &\scriptsize{Crystal Barrel}~\cite{Amsler_1995}\\
$\bar{p}d \to \eta n $& 3.19 $\pm 0.48 $ &$\times 10^{-6}$ & \scriptsize{Crystal Barrel}~\cite{Amsler_1995}\\
$\bar{p}d \to \omega n$ & 22.8 $\pm 4.1$ & $\times 10^{-6}$ & \scriptsize{Crystal Barrel}~\cite{Amsler_1995}\\
$\bar{p}d \to \eta' n$ & 8.2$\pm 3.4 $ &$\times 10^{-6}$ &  \scriptsize{Crystal Barrel}~\cite{Amsler_1995}\\
$\bar{p}d \to \phi n$ & 3.56 $\pm 0.20^{+0.2}_{-0.1} $ &$\times 10^{-6}$ & \scriptsize{OBELIX (in gas)}~\cite{Gorchakov_2002}\\
$\bar{p}d \to \rho^-p$ & 2.9 $\pm 0.6 $ &$\times 10^{-5}$ & \scriptsize{OBELIX (in gas)}~\cite{Ableev_1994}\\
$\bar{p}d \to \pi^-\Delta^+(\to\pi^0p)$ & 1.01 $\pm 0.08 $ &$\times 10^{-5}$ & \scriptsize{OBELIX (in gas)}~\cite{Denisov_1999}\\
$\bar{p}d \to \pi^0\Delta^0(\to\pi^-p)$ & 1.12 $\pm 0.20 $ &$\times 10^{-5}$ & \scriptsize{OBELIX (in gas)}~\cite{Denisov_1999}\\
$\bar{p}d \to \pi^0\Delta^0(\to\pi^0n)$ & 2.21 $\pm 0.24 $ &$\times 10^{-5}$ & \scriptsize{Crystal Barrel}~\cite{Amsler_B352_1995}\\
$\bar{p}d \to \Sigma^0 K^0$ & 2.15 $\pm 0.45 $ &$\times 10^{-6}$ & \scriptsize{Crystal Barrel}~\cite{Abele_1999}\\
$\bar{p}d \to \Lambda K^0 $& 2.35 $\pm 0.45 $ &$\times 10^{-6}$ & \scriptsize{Crystal Barrel}~\cite{Abele_1999}\\
\hline
\end{tabular}
\end{center}
\caption{Branching ratios of Pontecorvo reactions measured at LEAR in liquid deuterium and in gas}
\label{tab:Pontecorvo}
\end{table}

Some observations challenged existing theoretical predictions, such as the $\phi n$ production rate violating the OZI rule. 
Ratios like R($\phi/\eta$) = 0.156 ± 0.029 echoed patterns seen in free nucleon annihilations, suggesting contributions from 
nucleon short-range dynamics.
For channels with open strangeness, such as \(\overline{p}d \rightarrow \Lambda K^0\) and \(\overline{p}d \rightarrow \Sigma^0 K^0\), the measured branching ratios contradicted predictions from simple rescattering models. Instead, experimental results aligned better with a phase-space-based approach.

These studies highlighted the importance of nuclear binding and short-range effects in determining final states. 
Future investigations may extend these ideas to systems involving three nucleons, potentially offering even more stringent tests of theoretical models.

\section{Pontecorvo Reactions in 3-Nucleon Targets}

In addition to the measurements of antiproton annihilation in deuterium, other Pontecorvo reactions of significant interest remain unexplored experimentally.

Using deuterium as a target, the employment of antineutrons as projectiles would provide an opportunity to investigate a series of channels analogous to those reported for antiprotons in Table~\ref{tab:Pontecorvo}. Another possibility could involve the use of a deuteron beam~\cite{Caravita_2024}. Although such a high-intensity beam is not currently available, it could, in the future, enable the observation of the Pontecorvo reactions
$\bar{d} + d \to \bar{p} + p \quad$ and $\quad \bar{d} + d \to \bar{n} + n$

However, a more realistic approach would be to consider reactions involving antiprotons and targets with three nucleons, such as:
\begin{equation}
\bar{p} + ^3\!\text{He} \to n + p \label{eq:eq1}
\end{equation}
\begin{equation}
\bar{p} + ^{3}\!\mathrm{H} \to n + n
\label{eq:eq2}
\end{equation}
Between these, the measurement of the latter reaction is challenging due to safety concerns related to the use of tritium as a target. Conversely, the former reaction is experimentally feasible, and this article proposes some methods for its implementation. An evident advantage compared to the study in deuterium is that the rates predicted for the reaction~\eqref{eq:eq1} by the rescattering model and the fireball model are very different, being on the order of $10^{-8} - 10^{-7}$ for the former and $10^{-6}$ for the latter. 
In Figure 1, the graph of reaction~\eqref{eq:eq1} according to the rescattering model and the rerrangement graph in the fireball model are shown.


\begin{figure}[htbp]
    \hspace{1.cm}   
    \begin{subfigure}[b]{0.35\textwidth}
        \centering
        \includegraphics[width=\linewidth]{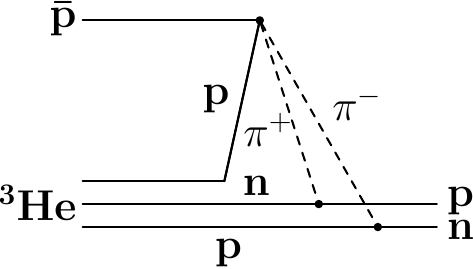}
        \label{fig:diagram1}
    \end{subfigure}
    \hspace{2.cm} 
    \begin{subfigure}[b]{0.35\textwidth}
        \centering
   
        \includegraphics[width=1\textwidth]{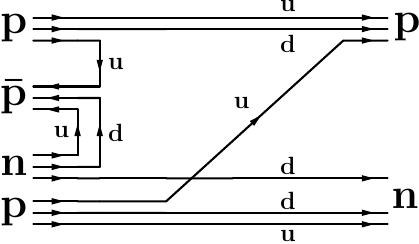}
        \label{fig:diagram2}
    \end{subfigure}
\vspace{-4mm} 
    \caption{Graph of reaction~\eqref{eq:eq1} according to the rescattering model (left) and the rerrangement graph in the fireball model (right).}
    \label{fig:two_diagrams}
\end{figure}

For completeness, the analogous reactions involving antineutrons are: $\quad \bar{n} + ^3\!\text{He} \to p + p \quad$ and $\quad \bar{n} + ^3\!\text{H} \to p + n$

\section{Antiproton Beams and a Simple Measurement Apparatus}
The study of Pontecorvo reactions has always been conducted using continuous antiproton beams and magnetic spectrometers as measurement apparatus. 
A continuous beam is essential because it allows the identification of rare Pontecorvo events among the much more frequent background of standard antiproton annihilations in the target, which must be rejected.  

In this article, we discuss both potential antiproton beams and an alternative measurement apparatus that could replace the traditional magnetic spectrometer, aiming for a more cost-effective solution. 

\subsection{Antiproton Beam}
Among the options considered, the antiproton beam from CERN’s ELENA facility offers a viable solution for studying Pontecorvo reactions, provided that the slow extraction mode is implemented~\cite{Gamba}. In its current operational mode, ELENA delivers pulsed beams with an antiproton energy of 100 keV. 
Each pulse provides approximately \(10^7\) antiprotons every two minutes to four experiments, with a bunch duration of a few hundred nanoseconds. If the same number of antiprotons were distributed continuously over the two-minute interval, one could achieve an effective rate of approximately \(10^5\) antiprotons per second.  

In the absence of ELENA's continuous beam mode, an alternative is the secondary beamline from the ASACUSA experiment~\cite{Kraxberger}, specifically designed for nuclear physics studies. The MUSASHI antiproton trap in ASACUSA can operate in two extraction modes: a pulsed extraction mode for the antihydrogen experiment and a DC extraction mode intended for nuclear studies. Slow extracted antiprotons from MUSASHI are transported to the annihilation target via the positron transfer line, spanning between the positron accumulator and the Cusp trap. Preliminary tests have successfully transported antiproton beams from MUSASHI along this line, achieving approximately 25,000 antiprotons per spill, corresponding to a transport efficiency of about 8\%~\cite{Kraxberger}
The efficiency is expected to improve, enabling the transport of 100,000 antiprotons per spill to the target. However, this intensity remains lower than that of ELENA's beam and presents challenges, including the requirement to accelerate the low-energy beam (hundreds of eV) to at least 100 keV to pass through the thin window separating the beamline from the gaseous \( ^3 \text{He} \)  target. Despite these limitations, ASACUSA's secondary line remains a posible alternative for studying Pontecorvo reactions in the absence of ELENA's continuous extraction capability.

\subsection{Target}
Simulations performed using Geant4 indicate that the maximum thickness of a plastic material window, such as Mylar, that can be nearly fully traversed by 100 keV antiprotons is approximately 1.2~\(\mu\text{m}\). Antiprotons exiting the window have an energy range of 0 to 30 keV. To stop these antiprotons in a gaseous  $^3$He target with a cylindrical geometry (10 cm height and 10 cm base diameter) aligned along the beam axis, the target density must correspond to a pressure of 200 mbar at room temperature. However, under these conditions, the Mylar window would be too fragile, as it would need to withstand a pressure differential of approximately 0.2 bar between the vacuum in the beamline and the gaseous target. To address this issue, the target should be cooled to cryogenic temperatures, reducing the pressure differential by two orders of magnitude and preventing the Mylar from breaking.

An alternative is to use a commercial silicon nitride (SiN) window with a thickness of 500 nm instead of Mylar. In this case, a 6 mm-wide SiN membrane would eliminate the need for cryogenic cooling, as similar membranes can withstand pressure differentials between 1 and 2 bar \cite{Silson}
.

\subsection{Measurement Apparatus}
Magnetic spectrometers were used in experiments to measure the branching ratios of Pontecorvo reactions in deuterium. These experiments investigated the interactions of antiprotons with nuclear matter, requiring precise instruments to detect and analyze the particles produced. 


The possibility of using a simpler and less expensive apparatus to detect the final state of the reaction~\eqref{eq:eq1} is considered here. 
The apparatus, in addition to detecting the proton, must include an efficient neutron detector and a veto system capable of rejecting background events, which consist of all the antiproton annihilations on quasi-free nucleons of the target. When an antiproton at rest interacts with a $^3\text{He}$ nucleus, it is initially captured to form an antiprotonic atom. After a radiative cascade to lower energy levels, the antiproton in almost all cases annihilates with a proton or neutron of the nucleus, contributing to the background of the measurement. This annihilation primarily produces pions, with a mean multiplicity of 5 (including 2 neutral pions), and less frequently strange particles such as kaons. The annihilation can also proceed through intermediate resonant states, which subsequently decay into pions and kaons. Table~\ref{tab:pbarp} provides the measured branching ratios for antiproton annihilation at rest in liquid hydrogen targets. 
Among the effects of the annihilation is the fragmentation of the nucleus into neutrons, protons, or deuterons.
\begin{table}[h!]
\centering
\begin{subtable}[t]{0.47\textwidth} 
\centering
\begin{tabular}{lll}
\toprule
Channel & BNL (\%) & CERN (\%) \\
\midrule
$\pi^+\pi^-$ & $0.37 \pm 0.03$ & $0.32 \pm 0.04$ \\
$\pi^+\pi^-\pi^0$ & $6.9 \pm 0.35$ & $7.3 \pm 0.9$ \\
$2\pi^+2\pi^-$ & $6.9 \pm 0.6$ & $5.8 \pm 0.3$ \\
$2\pi^+2\pi^-\pi^0$ & $19.6 \pm 0.7$ & $18.7 \pm 0.9$ \\
$3\pi^+3\pi^-$ & $2.1 \pm 0.2$ & $1.9 \pm 0.2$ \\
\bottomrule
\end{tabular}
\end{subtable}%
\hfill
\begin{subtable}[t]{0.50\textwidth} 
\centering
\begin{tabular}{lll}
\toprule
Channel & BNL (\%) & CERN (\%) \\
\midrule
$3\pi^+3\pi^-\pi^0$ & $1.9 \pm 0.2$ & $1.6 \pm 0.2$ \\
$n\pi^0, n > 1$ & $4.1 \pm 0.4$ & $3.3 \pm 0.2$ \\
$\pi^+\pi^-n\pi^0, n > 1$ & $35.8 \pm 0.8$ & $34.5 \pm 1.2$ \\
$2\pi^+2\pi^-n\pi^0, n > 1$ & $20.8 \pm 0.7$ & $21.3 \pm 1.1$ \\
$3\pi^+3\pi^-n\pi^0, n > 1$ & $0.3 \pm 0.1$ & $0.3 \pm 0.1$ \\
\bottomrule
\end{tabular}
\end{subtable}
\caption{Branching ratios for antiproton annihilation into pions in liquid hydrogen (from \cite{Lu1995}).}
\label{tab:pbarp}
\end{table}

To enable the measurement system to distinguish these events, it must exclude all cases involving neutral particles other than the fast neutron, particularly gamma rays arising from the decay of neutral pions formed during the annihilation of antiprotons on quasi-free nucleons in the target.
Conversely, to identify the reaction~\eqref{eq:eq1}, its kinematics must be exploited, where the two nucleons are emitted back-to-back with a kinetic energy of approximately 1 GeV.

A preliminary design of the apparatus, illustrated in Figure 2, was developed using Monte Carlo simulations performed with GEANT4. It features an approximately cubic structure with a side length of less than 1.5 meters and a geometric coverage of about 2/3 of the total solid angle.
\begin{figure}
\includegraphics[width=.9\textwidth, trim={0cm 4.5cm 0cm 4cm}, clip]{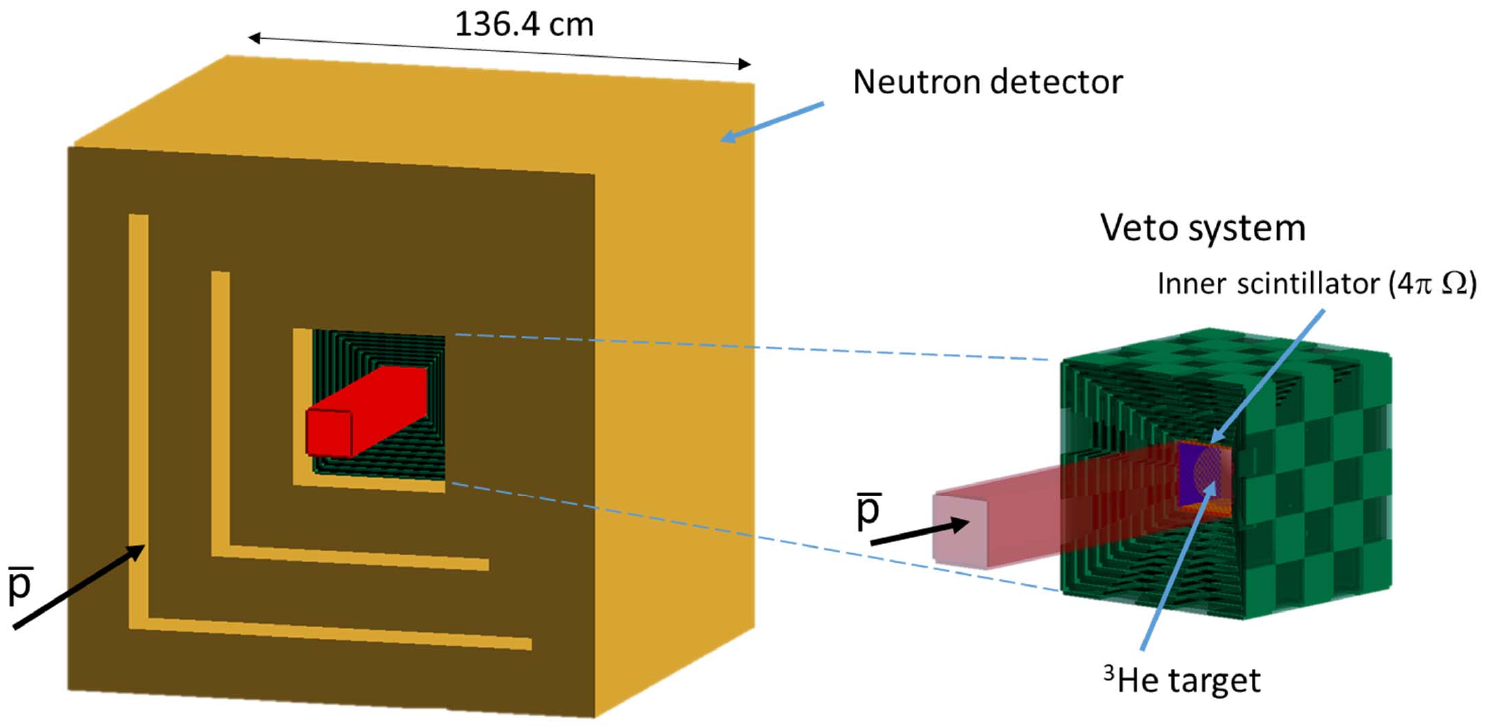}
\caption{Illustration of the proposed apparatus.}
\label{fig2}
\end{figure}

At the center lies the  $^3\text{He}$ target, with linear dimensions of 10 cm, housed in an aluminum container with a thickness of 0.32 mm. This thickness was chosen to allow spectator protons, emitted in background events with energies of a few tens of MeV, to pass through and be detected, thus improving the efficiency of background rejection. Surrounding the target are the veto system and, further outward, the neutron detector, both based on plastic scintillators (PVT).

The veto system covers four faces of the cubic structure, leaving uncovered the face from which the antiprotons enter and the opposite face. An exception is the innermost scintillator layer, which also includes the cubic face opposite to the antiproton beam direction. For this innermost layer, the faces parallel to the beam are extended with scintillators toward the antiproton source, achieving nearly \( 4\pi \) str solid angle coverage relative to the apparatus center and allowing more effective determination of the charged multiplicity of the annihilation products.
The veto system consists of a sandwich structure with 11 
layers of plastic scintillators, each 6 mm thick, alternating with 10 layers of lead, also 6 mm thick, in a configuration similar to a sampling calorimeter. The lead serves to promote the conversion of gamma rays, originating from the decay of neutral pions, into charged particles easily detectable by the veto scintillators.

The neutron detector also covers four faces and consists of three layers, each 15 cm thick. The total thickness of 45 cm represents a compromise between containing size and cost while maintaining adequate neutron detection efficiency. According to GEANT4 simulations, a plastic scintillator of this thickness can detect approximately 45\% of neutrons with 1 GeV energy passing through it perpendicularly.

The segmentation of the scintillators in both the veto system and the neutron detector was designed to ensure high spatial resolution. This optimizes the discrimination of useful events, effectively reducing background noise. The innermost scintillator of the veto system is divided into 20 × 20 cells for each of the five faces, with cell dimensions of \( 0.5 \times 0.5 \times 0.6 \, \text{cm}^3 \) per cell. The four extensions toward the antiproton source are 59 cm long and do not require segmentation.

The outermost veto scintillators, in contrast, have a coarser segmentation with 5 × 5 cells per face. The total number of readout channels is 9672. 

\subsection{Event selection}
To identify events associated with the Pontecorvo reaction~\eqref{eq:eq1}, an event selection based on event topology was developed. This requires that only a single hit be recorded in the innermost layer of the veto system, considered as the signal of the proton emitted during the reaction. 
This criterion rejects all background events with a measured charge multiplicity greater than one. 
Additionally, to distinguish the fast proton produced in the reaction~\eqref{eq:eq1} from the slow spectator proton typical of background events, the fast proton's signal must be accompanied by at least one hit recorded in the neutron detector along the charged particle path.
Furthermore, a signal is required in the neutron detector in the direction opposite to the proton impact. This signal is expected to originate from the neutron produced in the Pontecorvo reaction~\eqref{eq:eq1}.

To ensure the exclusion of background events, no signals should be detected in the veto system far from the proton track.
This combination of criteria allows for the optimization of the identification of events of interest, significantly reducing background contributions and ensuring more reliable data acquisition.

\section{Simulation Results}
Monte Carlo simulations with GEANT4 (version 10) evaluated the detection efficiency of the Pontecorvo reaction and background rejection. 
A custom class generated reaction~\eqref{eq:eq1}, while built-in functionalities handled background events. Although GEANT4 approximates antiproton annihilation products~\cite{Amsler_2024}, its FTFP\_BERT\_HP + STD + HP Physics List was used due to a lack of experimental data. A 500 keV detection threshold was applied to all detectors, with events generated at the target's center.
By applying the event selection described above, the detection efficiency for reaction~\eqref{eq:eq1} was found to be 11\%. The background rejection power was on the order of \(10^8\).
A similar result is also achieved for the reaction~\eqref{eq:eq2}.
Assuming that the production rate of Pontecorvo reaction is \(10^{-6}\), consistent with predictions from the fireball model, the signal-to-noise ratio is 10.
With these results, if a continuous beam from ELENA were available, only 1 spill (2 minutes) would be necessary to observe a single Pontecorvo event. In contrast, using the ASACUSA secondary line, approximately 90 spills, equivalent to 3 hours of data acquisition, would be required. It is also worth noting that acquiring additional data on antiproton annihilation in hydrogen would be valuable for further characterizing and reducing background contributions.

\section{Conclusions}
This study provides an overview of  Pontecorvo reactions, which are useful for investigating short-distance nucleon dynamics and antiproton annihilation mechanisms within nuclei.  Facilities such as CERN’s ELENA, provided it is modified to allow a slow extraction mode, and the ASACUSA beamline offer viable options for conducting these experiments, particularly with three-nucleon systems like \( ^3\text{He} \) and \( ^3\text{H} \).
The proposed experimental setup, utilizing plastic scintillators and veto systems, offers an alternative to traditional magnetic spectrometers. Simulations have shown that such setups can achieve reasonable detection efficiency and effective background rejection, making them suitable for the study of these rare processes.

\end{document}